\documentstyle[preprint,12pt,aps,epsf]{revtex}
\begin{document}

\title{First evidence for charge ordering in NaV$_2$O$_5$ from Raman spectroscopy 
}
\author{M.J.Konstantinovi\'c $^{a}$, Z.V.Popovi\'c $^{b}$, A.N.Vasil'ev$^{c}$,
M.Isobe$^{d}$ and Y.Ueda$^{d}$}
\address{$^a$ Max-Planck-Institut f\"ur
    Festk\"orperforschung,Heisenbergstr. 1,D-70569 Stuttgart, Germany}
\address{ $^b$ VSM-lab., Department of Physics, K.U.Leuven, Celestijnenlaan 200 
D,
Leuven 3001, Belgium}
\address{ $^c$ Low Temperature Physics Department, Moscow State
University, 119899 Moscow, Russia } 
\address{ $^d$ Institute for Solid State Physics, The
University of Tokio, 7-22-1 Roppongi, Minato-ku, Tokio 106, Japan}
\maketitle
\begin{abstract}
We argue on the basis of symmetry selection rules and Raman
scattering spectra that NaV$_2$O$_5$ undergoes a charge ordering phase
transition at T$_c$=34 K.  Such a transition is characterized by the
redistribution of the charges at the phase transition and corresponds
to the change of the vanadium ions, from uniform V$^{4.5+}$ to two
different V$^{4+}$ and V$^{5+}$ states.  In the low temperature phase the V$^{4+}$
ions are forming a "zig-zag" ladder structure along the {\bf b}-axis,
consistent with the symmetry of the P2/b space group.
\end{abstract}

keywords: Raman scattering; NaV$_2$O$_5$; Charge ordering;

NaV$_2$O$_5$ is reported to be the second inorganic spin-Peierls (SP)
compound \cite {a1} following CuGeO$_3$, with a phase transition temperature
T$_c$=34 K.  However, the following recent experimental findings suggest
 a character
for the phase transition of NaV$_2$O$_5$ rather different from CuGeO$_3$:

a) The large value of $2\Delta/k_BT_c=6$ shows strong deviation from
the results of the weak-coupling mean field theory of SP transitions
\cite{a1}. 

b) A strong thermal conductivity anomaly \cite {a2}, more
pronounced than that observed in CuGeO$_3$.  

c) T$_c$ is only weakly
dependent on an applied magnetic field \cite{a1}. 

 NaV$_2$O$_5$ is believed to be the
perfect realization of a quarter-filled low-dimensional system, with a
possible charge ordering into a geometrical configuration of the
charges with a spin-excitation gap \cite{a2,a3,a4,a5,a6,a7,a8}. 
 Such a scenario is supported by
structural analysis \cite{a3}, nuclear magnetic resonance (NMR) \cite{a4}, and
Raman scattering experiments \cite{a5}.  These studies showed that the V-ions 
are
equivalent at room temperature, and in an intermediate
 valence state V$^{4.5+}$.  On the
other hand, such a situation seems to disagree with the excellent fit
of a Bonner-Fisher curve to the magnetic susceptibility \cite{a1}.  An
explanation of these facts was proposed by Horsch and Mack \cite{a6} who
showed that due to the strong electron correlations the formation of
the Heisenberg antiferromagnetic chains are still possible even
without charge modulation. 

Low temperature NMR spectra \cite{a4} suggested
that the V-ions change their valence from uniform to alternating V$^{4+}$
and V$^{5+}$, thus
 giving the first direct evidence for the charge ordering
scenario.  So far, two different models for the electronic ordering
have been discussed.  One is described as "in-line" \cite{a7}, with V$^{4+}$ ions
arranged along the legs of the ladders ({\bf b}-axis) whereas the other
consists of V$^{4+}$ ions arranged in a "zig-zag" manner along the {\bf b}-axis
\cite{a8,a9}.  The first one requires an additional transition, such as an
ordinary SP transition, for the opening of the spin-gap.  For the
second one it is argued that the spin-gap appears due to "zig-zag"
ordering itself \cite{a9}.  However, in both cases, the driving mechanism
for the transition as well as the crystal symmetry have not been
elucidated. 

In the Raman spectra the charge ordering manifests itself through
the appearance of the new Raman active modes whose origin is still not
understood \cite{a10,a11,a12}.  In general, the full understanding of the
vibrational Raman spectra requires knowledge of the crystal structure.
In the case of NaV$_2$O$_5$ this information is not available (see note 
at end).  Still, in
some cases it is possible to perform the analysis of the crystal
symmetry starting from Raman spectra.  Here, we will show that such
analysis, with the help of IR spectra, allow us to make a plausible 
conjecture
the low-temperature space group of NaV$_2$O$_5$. 

The measurements were
performed on single crystals, with a size of approximately 1x3x1 mm$^3$ in
the {\bf a}, {\bf b}, and {\bf c} axis, respectively, prepared as described in 
\cite{a1}.  As
exciting source we used lines from Ar$^+$ and Kr$^+$ ion lasers.  The beam,
with an average power of 5 mW, was focused with macroscopic optics on
the (001), (010) and (100) surfaces of a crystal.  The spectra were
measured in back-scattering geometry using a DILOR triple
monochromator equipped with a CCD camera.  NaV$_2$O$_5$ crystallizes above
T$_c$ in the orthorombic centrosymmetric space group Pmmn (D$_{2h}^{13}$), 
with
two molecules in the unit cell of a size:  a=1.1318 nm, b=0.3611nm and
c=0.4797 nm.  Each vanadium atom, in the intermediate valence state 4.5+, is
surrounded by five oxygen atoms forming VO$_5$ pyramids.  These pyramids
are mutually connected via common edges and corners to form layers in
the (ab) plane.  The Na atoms are situated between these layers as
intercalants, see Fig. \ref{fig1}.  The structure of NaV$_2$O$_5$ can also be
described as an array of parallel ladders coupled in a trellis
lattice. 

A factor group analysis (FGA) yields the irreducible
representations of the vibrational modes in NaV$_2$O$_5$ \cite{a5}:

$\Gamma_{opt}$=8A$_g$(aa,bb,cc)+3B$_{1g}$(ab)+8B$_{2g}$(ac)+5B$_{3g}$(bc)+
7B$_{1u}$({\bf E}$||${\bf c})+4B$_{2u}$({\bf E}$||${\bf b})+7B$_{3u}$({\bf 
E}$||${\bf a})

Our previously reported room temperature Raman and IR
spectra \cite{a5} confirmed two important consequences of this symmetry
analysis:  1. The mutual exclusion of the Raman and IR active modes
(corresponding to the existence of a center of inversion).  2.  Good
agreement between observed and predicted number of modes for each
polarized configuration. 

In Fig. \ref{fig2} we present the first complete Raman spectra of a
NaV$_2$O$_5$ single crystal at room temperature and at T=10 K in all 
inequivalent polarized configurations. Some new modes that become
active below the phase transition temperature T$_c$=34 K, are 
appropriately marked in
the graph. The full list of Raman actuve modes, below and above the
phase transition temperature, together with the IR data of Damascelli
et al. \cite{a13} is given in Table 1. The low-temperature modes are
 assigned either to magnetic exitations (66, 106 and 
132 $cm^{-1}$, magnetic bound states) or to phonon folded modes 
due to the supersymmetry \cite{a11}. The origin of the continuum-like 
features
 (the most intense ones
observed in aa and ab spectra) has not been elucidated and
 will not be discussed here. 
However, it is interesting that most of the low-temperaure modes appear
 either in the
(aa), (bb), (cc) and (ab) - polarized configurations or in the (ac)
and (bc) configurations. The frequencies of these modes are marked
 in the shaded regions of 
Table 1. This is a rather important property of the
crystal symmetry at low-temperatures.  The second piece of information
for  determining the possible crystal symmetry comes from the low 
temperature IR
spectra.  Damascelli et al. \cite{a13} observed many equal or nearly equal
frequencies for  {\bf E}$||${\bf a} and 
{\bf E}$||${\bf b} (see Table 1). Moreover,
we did not find any modes with the same energy in the Raman and IR
spectra. 
 These experimental facts lead to the following conclusions
for the low-temperature crystal symmetry.  1.  There exists a center
of inversion.  2.  The crystal symmetry must be lower than
orthorhombic.  At this point we can already conjecture that the factor
group (i.e. the point group) symmetry is C$_{2h}$.

Let us now look for
the possible charge ordering patterns which are consistent with the
symmetry properties described above.  The first one we considered, the
"in-line" charge ordering pattern, is ruled out immediately since it
does not have a center of inversion as a symmetry operation.  For
"zigzag" charge ordering we found two possible patterns.  These
patterns, together with their symmetry operations are shown in
Fig. \ref{fig3} and Fig. \ref{fig4} (shaded VO$_{5}$ pyramids represent the 
V$^{4+}$ valence states).
The first one (we refer to it as "zz1") has four
symmetry operations:   Identity,   Center of inversion, a  C$_2$
twofold rotation along the {\bf c}-axis and a  c-glide plane,
 see Fig. \ref{fig3}.
According to \cite{a14}
the space group with these properties is P2/b (C$_{2h}^4$).  The other
zig-zag pattern called "zz2" (see Fig. \ref{fig4}) has, 
instead of the rotation C$_2$,
a screw axis C$_2$ along the {\bf a}-axis.  In this "zz2" case the
corresponding space group is P2$_1$/a (C$_{2h}^5$).  In both cases the unit
cell is twice larger than for the room temperature phase and
comprises four molecules.  The site symmetry of all atoms is C$_1$ since
only C$_1$ can accommodate four atoms in both C$_{2h}^4$ and C$_{2h}^5$. 
Then, the
FGA \cite{a15} yields:

(C$_1$): $\Gamma$=3A$_g$+3A$_u$+3B$_g$+3B$_u$

Taking each of these
representations 8 times (4x8 is the total number of atoms in the unit
cell) and subtracting the three acoustic modes A$_u$+2B$_u$ we obtain the
irreducible representations of the NaV$_2$O$_5$ vibrational modes for either
the C$_{2h}^4$ or the C$_{2h}^5$ space groups.

$\Gamma_{opt}$=24A$_g$(aa,bb,cc,ab)+ 24B$_g$(ac,bc)+23A$_u$({\bf E}$||${\bf c})+
22B$_u$({\bf E}$||${\bf a} and {\bf E}$||${\bf b});  P2/b (C$_{2h}^4$)

$\Gamma_{opt}$=24A$_g$(aa,bb,cc,bc)+ 24B$_g$(ab,ac)+23A$_u$({\bf E}$||${\bf a})+
22B$_u$({\bf E}$||${\bf b} and {\bf E}$||${\bf c});  P2$_1$/a (C$_{2h}^5$)

Having in mind all facts discussed above, we
can conclude that the proper space group of NaV$_2$O$_5$ in the low
temperature phase is P2/b (C$_{2h}^4$).  This space group, with a unit cell
shown in Fig. \ref{fig3} explains the doubling of the unit cell along {\bf a}
 and {\bf b}
directions observed in the x-ray data \cite{a16}.  Quadrupling of the
low-temperature unit cell along the {\bf c}-axis appears because of the four
possible zz1 arrangements in the (ab) plane.  In fact, we derive the
irreducible representations of the NaV$_2$O$_5$ vibrational modes of
the C$_{2h}^4$
space group by assuming only four molecules in the unit cell (one
layer).  By taking into account all four layers, 16 molecules, the
number of the vibrational modes has to be multiplied by four and the 
corresponding space group is B2/b (C$_{2h}^6$).
However, the number of modes observed in our Raman spectra indicate
the dominance of the layer symmetry.  Still, the proper monoclinic
unit cell does not correspond to the one depicted in Fig. \ref{fig3} and one
needs to introduce additional distortions of the cell.  Most likely,
these distortions correspond to vanadium V$^{4+}$ ion displacements along
the ab direction because of the Coulomb repulsion.  Such atomic
displacements are indicated in Fig. \ref{fig3} by arrows.
We shall discuss later the
relevance of such distortions to the magnetic properties of NaV$_2$O$_5$.

The "zig-zag" charge ordering configuration naturally imposes the
antiferromagnetic "dimer" pattern along the ab direction \cite{a8}.
However, Hartree-Fock calculations
showed that the "zig-zag" charge ordering in NaV$_2$O$_5$ is driven by
intersite Coulomb interaction, with the result not being sensitive to
the exact value of the hopping integral t$_{xy}$ \cite{a8}.
  Therefore the formation
of the "dimers" along the ab direction is not obvious.  We also know
from the work of Gros and Valent\'\i \cite{a17} that "zig-zag" charge ordering
explains the splitting of the magnon dispersion branch along the
{\bf a}-axis \cite{a18}.  Unfortunatelly, in their model the modulation of the
exchange integral along the {\bf b}-axis, which is essential for the opening
of the spin gap, is introduced by hand, without detailed explanation.
In any case, the spin-gap behavior in the magnetic susceptibility
measurements can be understood within the framework of the
spin-singlet formation, but with the corresponding state not being the
SP state.  Therefore, the comparison of $2\Delta/k_BT_c$ to the corresponding
values for other SP compounds is irrelevant.  Our analysis of the
NaV$_2$O$_5$ crystal structure suggests that the crystal space group symmetry
P2/b, together with the distortion pattern shown in Fig. \ref{fig3}, leads to
the alternation of the exchange integral along the {\bf b}-axis and directly
supports the model of Gros and Valent\'\i{} if one assumes small interladder
exchange coupling ($J_{1}$ and $J_{2}$ are exchange constants along the 
ladders while $J_{3}$ represents the interladder exchange).  
In this case "dimers" are formed in the
{\bf b}-direction and magnetic dynamics should be analyzed using a 1D
dimerized Hamiltonian with frustration.  If one assumes the opposite
case for the distortions, i.e., movement of V$^{4+}$ towards each other,
which is also consistent with the proper monoclinic unit cell, one is
led to the solution of Seo and Fukuyama \cite{a8},
 with large t$_{xy}$.  In order to
resolve this puzzle precise low temperature x-ray measurements are
needed. 

In conclusion, on the basis of symmetry selection rules and
Raman scattering spectra we argue that NaV$_2$O$_5$ undergoes a charge
ordering phase transition at T$_c$=34 K.  Such a transition is
characterized by the redistribution of the charges at the phase
transition and corresponds to the change of the valences of the
vanadiun ions from uniform V$^{4.5+}$ to two different V$^{4+}$ and V$^{5+}$ 
states.
In the low temperature phase the V$^{4+}$ ions are forming a "zig-zag"
ladder structure along the {\bf b}-axis, consistent with symmetry
properties of the P2/b (or B2/b - four layers) space group.
Moreover, this crystal symmetry
naturally imposes displacements of vanadium V$^{4+}$ ions along the ab
direction, which lead to the alternation of the exchange integral
along either the {\bf b}-axis or the ab direction.

Note added:  After completion of this work an analysis of the
low-temperature NaV$_2$O$_5$ crystal structure, using synchrotron radiation
x-ray diffraction, appeared in \cite{a19}.  The authores
 found that a centrosymmetric
average structure persists down to 15 K, with a modulation pattern
described by an acentric Fmm2 superstructure.  However, the conclusions
are inconsistent with our optical data which clearly show the
existence of Raman tensors that are characteristic for the
monoclinic symmetry.  The explanation may lie in the fact that they
also found a few reflections in x-ray spectra obeying C centering
which they did not consider in details.

It is a pleasure to thank M.Cardona, M.Cuoco,
A.Damascelli and P.H.M van Loosdrecht for helpful discusions.  MJK
thanks the Roman Herzog-AvH foundation for financial support.

\begin{figure}
\caption {Schematic representation of the room
temperature NaV$_2$O$_5$ crystal structure in the a) (010), c) (100), d)
(001) planes.  b) The symmetry operations represented in the (001)
plane.} 
\label{fig1}
\end{figure}
\begin{figure}
\caption
{Raman spectra of NaV$_2$O$_5$ at room temperature (dashed line) and at T=10
K (full line) in (aa) , (bb), (cc), (ab), (bc), and (ac) polarized
configurations, measured with 514.5 nm laser line.  The new modes that
appear below the phase transition temperature are marked with numbers
in rectangle.}
\label{fig2}
\end{figure} 
\begin{figure}
\caption 
{Schematic representation of the low temperature NaV$_2$O$_5$
"zig-zag (zz1)" crystal structure in the (001) plane.  The arrows
indicate the displacements of V$^{4+}$ ions.  The full/dashed lines
represent intra/inter ladder exchange interactions. }
\label{fig3}
\end{figure} 
\begin{figure}
\caption
{Schematic representation of the low temperature NaV$_2$O$_5$
"zig-zag (zz2)" crystal structure in the a) (001) and b) (100) planes.}
\label{fig4}
\end{figure} 

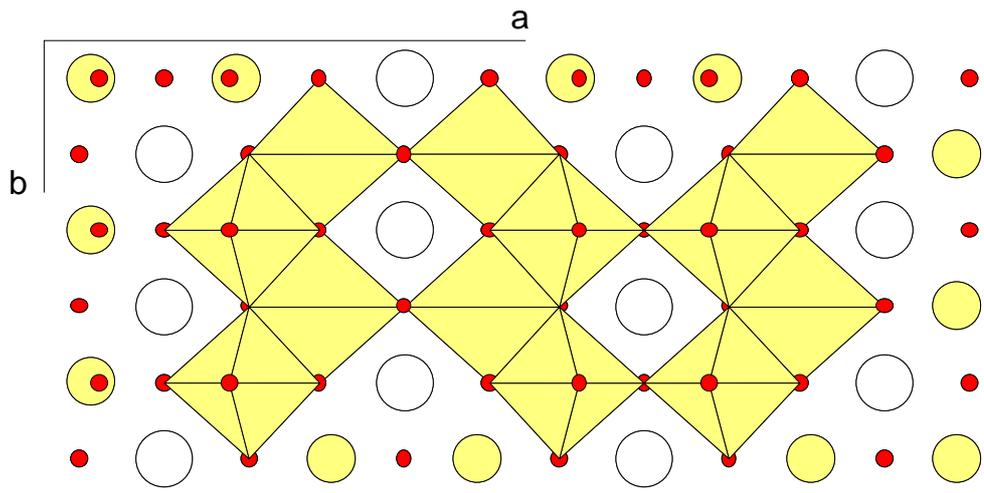
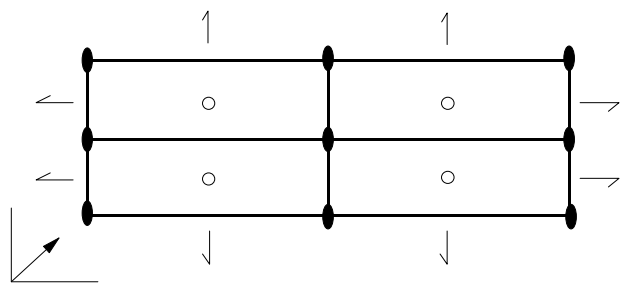
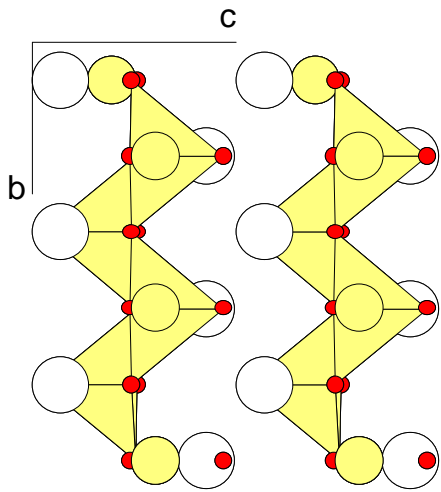
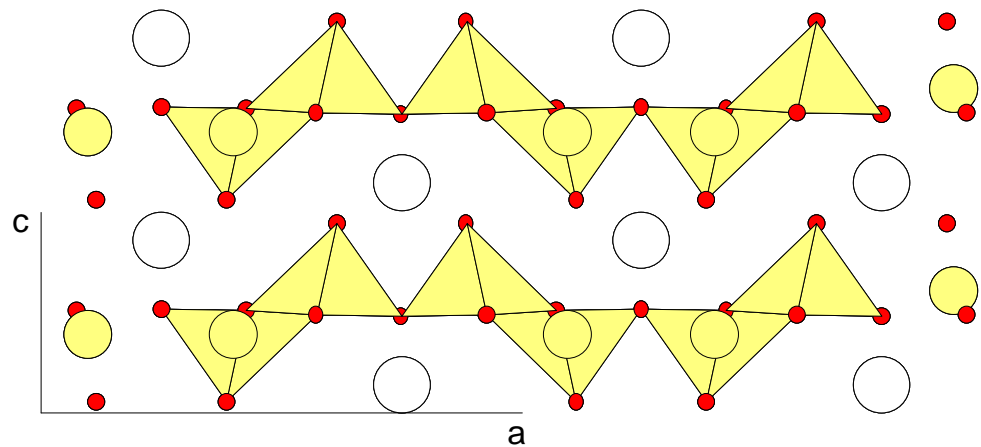
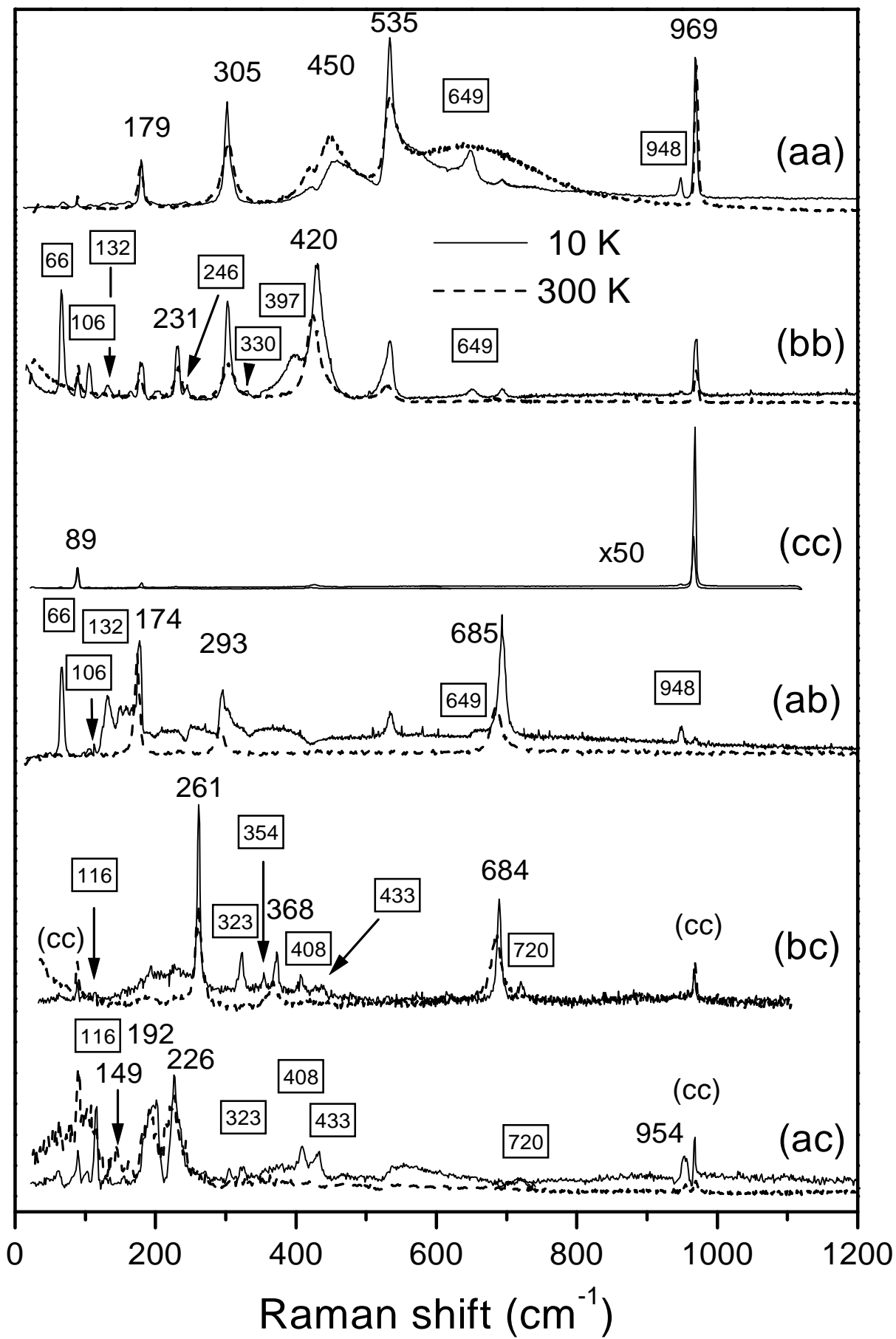
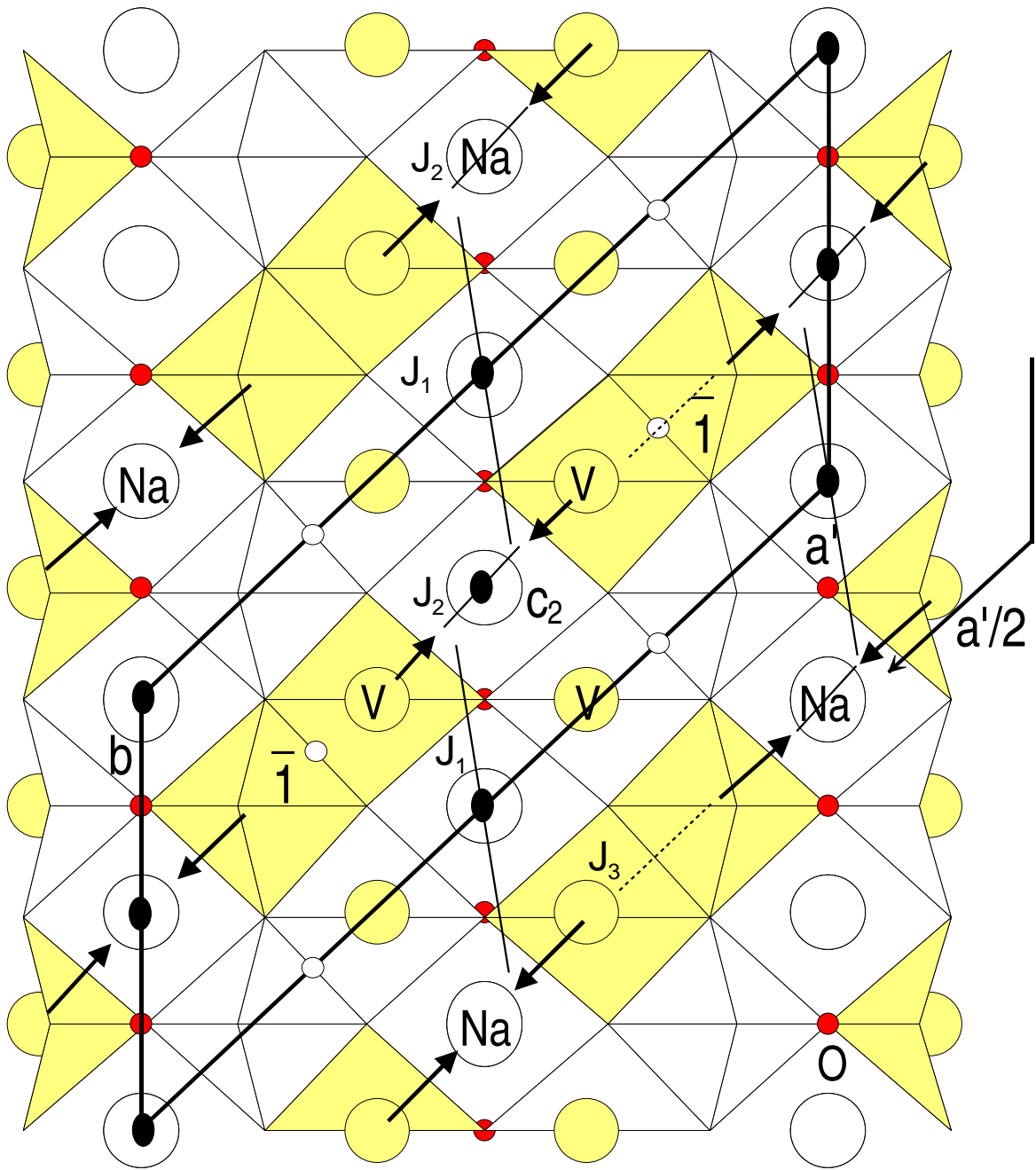
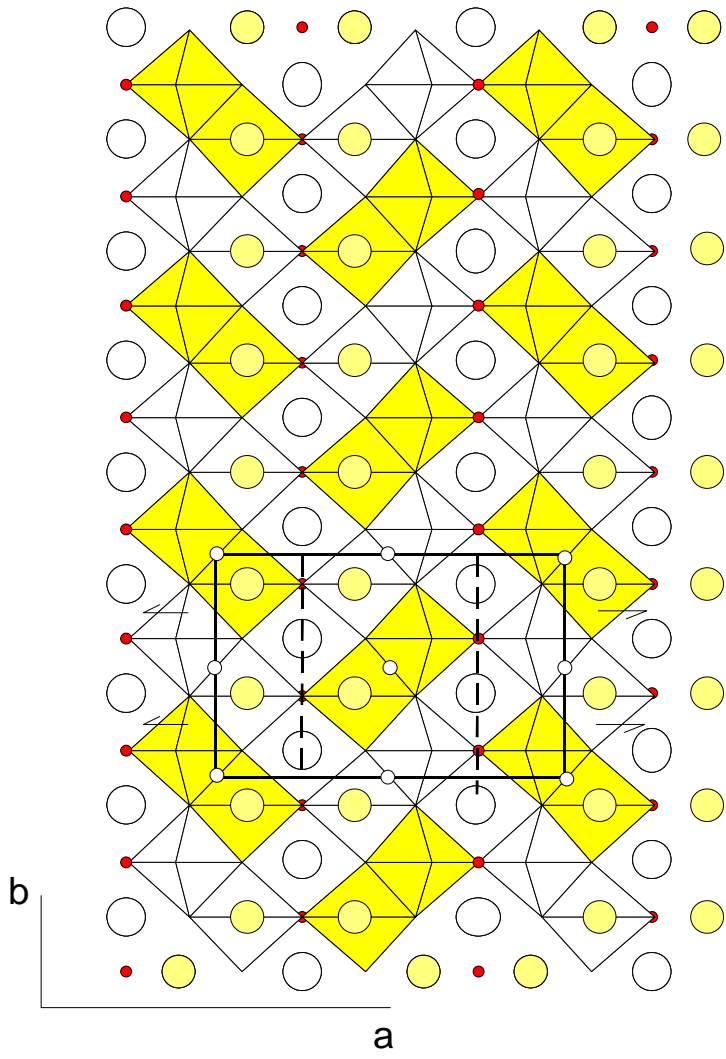
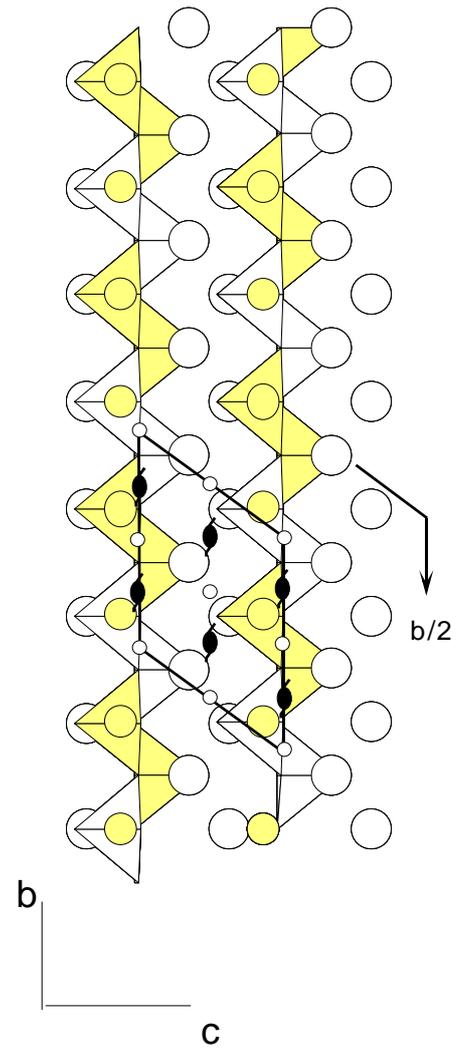
\end{document}